\documentstyle[12pt]{article}
\begin{document}
\bibliographystyle{unsrt}
\newcommand{\bra}[1]{\left < \halfthin #1 \right |\halfthin}
\newcommand{\ket}[1]{\left | \halfthin #1 \halfthin \right >}
\newcommand{\be}{\begin{equation}}
\newcommand{\ee}{\end{equation}}
\newcommand{\vsig}{\mbox {\boldmath $\sigma$\unboldmath}}
\newcommand{\vep}{\mbox {\boldmath $\epsilon$\unboldmath}}
\newcommand{\fn}{\frac 1{E_N+M_N}}
\newcommand{\fs}{\frac 1{E_f+M_f}}

\title{\bf  An Unified Approach To Pseudo Scalar Meson
Photoproductions Off Nucleons In The Quark Model}

\author{Zhenping Li$^1$\thanks{E-mail; ZPLI@ibm320h.phy.pku.edu.cn}, 
Hongxing Ye$^1$ and Minghui Lu$^2$
\\
$^1$Physics Department, Peking University \\
Beijing 100871, P.R.China\\
$^2$ Institute of High Energy Physics, Beijing 100039, P.R.China}

\maketitle

\begin{abstract}
An unified approach to the pseudo scalar meson ($\pi$, $\eta$, and $K$)
 photoproduction off nucleons are presented.  
It begins with the low energy QCD Lagrangian, and the resonances in 
the s- and u- channels are treated in the framework of the 
quark model.  The duality hypothesis is imposed to limit the number of 
the t-channel exchanges.  The CGLN amplitudes for each reaction 
are evaluated, which include both proton and neutron targets.  The 
important role by the S-wave resonances in the second resonance region
is discussed, it is particularly important for the $K$, $\eta$ and 
$\eta^\prime$ photoproductions.
\end{abstract}

\newpage
\subsection*{\bf 1. Introduction}

Recently, there have been considerable interests to study
the meson photoproductions off nucleons.  The data
from ELSA in the kaon\cite{bock94} and $\eta$\cite{price} productions,
from MAMI\cite{krusch} and BATES\cite{dytman} in the
$\eta$ production have been published.
Further experiments have also been planned at the Continuous Electron 
Beam Accelerator Facility (CEBAF)\cite{shumacker} and other electron 
accelerator facilities,  which will provide a complete set of data in
$\pi$, K, and $\eta$ photoproductions with much better energy and 
angular resolutions.  This provides us a golden opportunity to study
the structure of baryon resonances and a chellange to understand the
reaction mechanism in terms of quantum Chromodynamics(QCD).

The theoretical investigations of meson photoproductions during 
the past 30 years have been concentrated in the isobaric models
\cite{thom,as90,as95,wjs,benn}, in which the 
Feynman diagrammatic techniques are used so that the transition
amplitudes are Lorentz invariant.  The recent investigations by
J-C David{\it et al.}\cite{as95} in the kaon photoproductions and
 by the RPI group\cite{muko} in the threshold region of the $\eta$ 
photoproduction have been quite successful in describing the 
available data.  Because the meson baryon interactions are treated
in the phenomenological level,  the isobaric models have no
explicit connection with QCD, and the number of parameters in these
models are generally related to the number of resonances that are
included in calculations.  Thus, it becomes increasingly important
to investigate the reaction mechanism in term of quarks and gluons
degrees of freedom.  Such a program has its genesis with the early 
work of Copley, Karl and Obryk\cite{cko} and Feynman, Kisslinger
 and Ravndal\cite{fkr} in the pion photoproduction,
 who provided the first clear evidence of underlying $SU(6)\otimes 
O(3)$ structure to the baryon spectrum. The following 
 calculations and discussions with the consistent treatment of
 the relativistic effects\cite{close} have not changed the
 conclusions of Refs. \cite{cko} and \cite{fkr} 
significantly.  These calculations in the framework of the quark
models have been limited on the transition amplitudes that are 
extracted from the photoproduction data by the phenomenological 
models.  The challenge is whether one could go one step 
further to confront the photoproduction data directly with the 
explicit quark and gluon degrees of freedom.  Such a step is
 by no means trivial, since it requires
that the transition amplitudes in the quark model  have 
correct off-shell
behavior, which are usually evaluated on shell.  More importantly, 
it also requires that the model with explicit quark and gluon degrees
of freedom give a good description of the contributions to the
photoproductions from the non-resonant background,  which are usually
used to evaluate the contributions from s-channel resonances.  
The low energy theorem in the threshold pion photoproduction is a 
crucial test in this regard, which the non-resonant contributions
dominate in the threshold region.  Our 
investigation\cite{zpli94} showed that the simple quark model is
no longer sufficient to recover the low energy theorem, and one has 
to rely on low energy QCD Lagrangian so that the meson baryons 
interaction is invariant under the chiral transformation.  Moreover,
we found substantial contributions from the S-wave resonances in the
second resonance region to the $E_{0+}$ amplitudes of the neutral
pion photoproductions.  This shows the importance of the consistent
treatment of both resonant and non-resonant contributions even in 
the threshold pion photoproductions.  We have 
extended it to the kaon\cite{zpli951} and $\eta$\cite{zpli952} 
photoproductions by combining the low energy QCD Lagrangian
 and the quark model, and the initial results showed very
 good agreements between the theory and experimental data 
 with far less parameters.
The purpose of this paper is to present a comprehensive and 
unified approach to the meson photoproductions based on the low 
energy QCD Lagrangian.  The duality hypothesis is also
imposed to limit the number of the t-channel exchanges, which 
was not done in our previous investigation\cite{zpli951}.  This 
reduces the number of free parameters even further, and in principle,
there is only one parameter for each isospin channel, such as
 $\alpha_{\eta NN}$ in the $\eta$ production or $\alpha_{KN\Lambda}$
and $\alpha_{KN\Sigma}$ for kaon productions.

The paper is organized as follows.  In the section 2, the 
theoretical framework is established in meson photoproductions 
starting from the low energy QCD Lagrangian.  The formalism 
in the chiral quark model is presented for the s and u channel
resonances in section 3. We shall show how the CGLN amplitudes 
for the s- and u- channel resonances are derived in the quark model.
 Although our approach start with the 
low energy QCD Lagrangian, it could also be extended to the 
heavy pseudoscalar meson photoproduction, such as the $\eta^\prime$
production,  as the meson quark coupling should also be either
pseudo scalar or pseudo vector for the $\eta^\prime$.  In section
4, we discuss some important features of the quark model approach 
to meson photoproductions.  In particular, the $S$-wave resonances
in the second resonance region play an important role in the 
threshold region of $K$, $\eta$ and $\eta^\prime$ photoproductions.
Finally, the conclusion will be given in section 5.

\subsection*{\bf 2. The Model}

To understand many of the successes of the nonrelativistic quark 
model, Manohar and Georgi proposed\cite{MANOHAR} the concept of
chiral quarks, which is described by the effective Lagrangian
\begin{equation}\label{1}
{\cal L}={\bar \psi} \left [ \gamma_{\mu} (i\partial^{\mu}+ V^\mu+\gamma_5
A^\mu)-m\right ] \psi + \dots
\end{equation}
where the vector and axial currents are
\begin{eqnarray}\label{2}
V_\mu=\frac 12\left (\xi^\dagger 
\partial_\mu\xi+\xi\partial_\mu\xi^\dagger\right ) ,
\nonumber \\
A_\mu=i\frac 12\left (\xi^\dagger \partial_{\mu} \xi -\xi\partial_{
\mu} 
\xi^\dagger\right ), 
\nonumber \\
\xi=e^{i\pi/f}
\end{eqnarray}
$f$ is a decay constant, the quark field $\psi$ in the $SU(3)$ symmetry
is 
\begin{equation}\label{222}
\psi=\left ( \begin{array}{c} \psi(u) \\ \psi(d) \\ \psi(s) \end{array}
\right ),
\end{equation}
and the field $\pi$ is a $3\otimes 3$ matrix;
\begin{equation}\label{3}
\pi=\left| \begin{array}{ccc} \frac 1{\sqrt {2}} \pi^0+\frac 1{\sqrt{6}}\eta 
& \pi^+ & K^+ \\ \pi^- & -\frac 1{\sqrt {2}}\pi^0+\frac 1{\sqrt {6}}\eta & 
K^0 \\ K^- & \bar {K}^0 &-\sqrt{\frac 23}\eta \end{array}\right|,
\end{equation}
in which the pseudoscalar mesons, $\pi$, $K$ and $\eta$, are treated
as Goldstone bosons so that the Lagrangian in Eq. \ref{1} are 
invariant under the chiral transformation. Starting from this chiral
Lagrangian, there are four components for the photoproductions of 
pseudoscalar mesons;
\begin{eqnarray}\label{4}
{\cal M}_{fi}=\langle N_f| H_{m,e}|N_i \rangle + 
\sum_j\bigg \{ \frac {\langle N_f|H_m |N_j\rangle 
\langle N_j |H_{e}|N_i\rangle }{E_i+\omega-E_j} \nonumber \\
 +\frac {\langle N_f|H_{e}|N_j\rangle \langle N_j|H_m
|N_i\rangle }{E_i-\omega_m-E_j}\bigg \}+{\cal M}_T
\end{eqnarray}
where $N_i(N_f)$ is the initial (final) state of the nucleon,
and $\omega (\omega_{m})$ represents the energy of incoming 
(outgoing) photons(mesons).  

The first term in 
Eq. \ref{4} is a seagull term, it is generated by the gauge 
transformation of the axial vector $A_{\mu}$ in the QCD Lagrangian.
The corresponding quark-photon-meson vertex is given by
\begin{equation}\label{5}
H_{m,e}=\sum_j \frac {e_m}{f_m}\phi_m {\bar \psi}_{j}(q_f) 
\gamma_{\mu}^j\gamma_{5}^j \psi_j(q_i)A^{\mu}({\bf k},{\bf r}_j),
\end{equation}
where $A^{\mu}({\bf k},{\bf r}_j)$ and $\phi_m$ are the electromagnetic 
and meson fields respectively. 
Notice that the seagull term in Eq. \ref{5}  is proportional to the 
charge $e_m$ of the outgoing mesons,  it does not contribute
to the productions of the charge neutral mesons. As it will be shown
later, this also leads to the forward peaking in differential 
cross sections for the charge meson production.

The second and the third terms are s- and u-channel contributions.  
There has been considerable information on the s- and u- channel 
resonances from $\pi$ N scattering as well as the pion 
photoproductions, and the transition properties of these resonances,
such as the electromagnetic transition as well as the meson decays, 
have been investigated extensively in the framework of the quark 
model.  Our task in meson photoproductions off nucleons is to combine
the electromagnetic transitions and the meson decays of these resonances
together, in particular those evaluated in Ref. \cite{simon},
and to express these transition amplitudes in terms of the
standard CGLN amplitudes\cite{cgln} so that the various experimental 
observables could be easily calculated\cite{tabakin}.   This has been
done for the proton target in the Kaon and the $\eta$ production,  
we will present the complete evaluation of the CGLN amplitudes for
the transitions from both proton and neutron targets to the resonances
below 2 GeV,  which corresponds to main quantum number in the harmonic 
oscillator wavefunction $n\le 2$ in $SU(6)\otimes O(3)$ symmetry limit.
The connection between the CGLN amplitudes for baryon resonances and
the helicity amplitudes, $A_{1/2}$ and $A_{3/2}$, in the electromagnetic
transition could be easily established,
this has been discussed extensively in Ref. \cite{zpli951}.  
For those resonances above 2 GeV, there is little information on their
properties,  thus they are treated as degenerate so that the 
contributions from the resonances with quantum number $n$ could be 
expressed in a compact form.  Generally, the contributions from 
those resonances with the largest spin for a given quantum number $n$
are the most important as the energy increases, this corresponds to 
spin $J= n+1/2$ for the processes $\gamma N \to K \Lambda$ and 
$\gamma N\to \eta N$ , and $J=n+3/2$ for the reactions 
$\gamma N\to K\Sigma$ and $\gamma N\to \pi N$. 

The contributions from the u-channel resonances are  divided into
two parts.  The first part is the contributions from the 
resonances with the quantum number $n=0$, which include the spin
1/2 resonances, such as the $\Lambda$, $\Sigma$ and the nucleon, and
the spin 3/2 resonances, such as the $\Sigma^*$ in kaon productions
and $\Delta(1232)$ resonance in $\pi$ productions.  Because of 
the mass splitting between spin 1/2 and 3/2 resonances with $n=0$
 are significant, they have to be treated separately.  
The transition amplitudes for these u-channel
resonances will also be written in terms of the CGLN amplitudes, 
which will be given in next section.  The second part comes from the
resonances with the quantum number $n\ge 1$.  
As the contributions  from the u-channel resonances 
are not sensitive to the precise mass positions,
they are treated as degenerate as well, so that the
contributions from these resonances could also be written in a compact
form, which is also in terms of the CGLN amplitudes.

The last term in Eq. \ref{4} is the t-channel charged meson exchange,  
it is proportional to the charge of outgoing mesons as well,  thus
it does not contribute to the process $\gamma N\to \eta N$. 
This term is required so that the total transition amplitude in Eq. 
\ref{4} is invariant under the gauge transformation\cite{bn95}.  
The other t-channel exchanges, such as the $K^*$ and $K1$ exchanges
 in the kaon 
production, which played an important role in Ref. \cite{as90,as95},
the $\rho$ and $\omega$ exchanges in the $\eta$ production are 
excluded due to the
constraint of the duality hypothesis.  This was not imposed
in our early investigation\cite{zpli951}
of the kaon photoproduction, in which the contribution from the $K^*$
exchange was included.   The duality hypothesis states that
the inclusion of the t-channel exchanges may lead to a double 
counting problem if a complete set of resonances is introduced
in the s- and u- channels.  Dolen, Horn and Schmid\cite{dolen}
found that the t-channel $\rho$ Reggie trajectories, which govern 
the asymptotic high energy behavior, can be extracted from 
s-channel $N^*$ resonances of their low energy model.  
This constraint has been applied to the kaon photoproduction 
by Williams, Ji, and Cotanch\cite{wjs}  in their quantum 
hadrondynamic approach. The problem in their approach is that 
the minimum set of the s- and u-channel resonances was used 
in the calculation so that the model space is severely truncated 
from a complete set of resonances to a few resonances,  in particular,
the resonances with higher spins that are important in higher energies
are neglected.  Thus, the theory becomes an effective theory, and the
duality constraint is less significant.  
On the other hand, the chiral quark model provides an ideal framework
to apply the duality constraint since every resonance could be included 
in principle and without additional parameters.
The explicit expression for the t-channel charged meson exchange
 is shown in the Appendix.

\subsection*{\bf 3. Formalism}

Before we present our chiral quark model approach, it is very useful to 
review some basic kinematic feature of meson photoproductions.  
The differential cross section for meson photoproductions in the center 
of mass frame  is
\begin{equation}\label{6}
\frac {d\sigma^{c.m.}}{d\Omega}=\frac {\alpha_e \alpha_m(E_N+M_N)(E_f+M_f)}
{4s(M_f+M_N)^2}\frac {|{\bf q}|}{|{\bf k}|} |{\cal M^\prime}_{fi}|^2
\end{equation}
where the factor $eg_A/f_m$ is removed from the transition matrix 
element ${\cal M^\prime}_{fi}$ so that it becomes dimensionless, and 
$\sqrt {s}=E_N+\omega_\gamma=E_f+\omega_m$ is the total energy in the c.m. 
frame.
The coupling constant $\alpha_m$ is related to the factor $g_A/f_m$ by the 
generalized Goldberg-Treiman relation\cite{cad90}, however, the quark mass 
effects lead to about 30 percent deviation from the measured value, while 
the Goldberg-Trieman relation is accurate within 5 percent for the 
pion couplings\cite{holstein}. Therefore, the coupling $\alpha_m$ will be 
treated as a free parameter for $K$ and $\eta$ productions 
at present stage.   

The transition matrix element ${\cal M^\prime}_{fi}$ is expressed in 
terms of the CGLN amplitude,
\begin{equation}\label{7}
{\cal M^\prime}_{fi}={\bf J \cdot \vep}
\end{equation}
where $\vep$ is the polarization vector of incoming photons, and the 
current $J$ is written as
\begin{equation}\label{8}
{\bf J}=f_1 \vsig+ if_2 \frac {(\vsig \cdot {\bf q})({\bf k}\times \vsig)}
{|{\bf q}| |{\bf k}|}+f_3\frac {\vsig \cdot {\bf k}}{|{\bf q}||{\bf k}| 
}{\bf q}+f_4\frac {\vsig \cdot {\bf q}}{{\bf q}^2}{\bf q}
\end{equation}
in the center mass frame, where $\vsig$ is the spin operator for the 
initial and final states with spin 1/2.  Therefore, the differential cross 
section in terms of the CGLN amplitude is\cite{tabakin}
\begin{eqnarray}\label{9}
|{\cal M}^\prime_{fi}|^2= & Re &\bigg \{ |f_1|^2+|f_2|^2-2\cos(\theta)
f_2f_1^*\nonumber 
\\ & + & \frac 
{\sin^2(\theta)}2 \left [ 
|f_3|^2+|f_4|^2+2f_4f^*_1+2f_3f_2^*+2\cos(\theta)f_4f_3^*\right ]
\bigg \}
\end{eqnarray}
where $\theta$ is the angle between the incoming photon momentum
 ${\bf k}$ 
and outgoing meson momentum ${\bf q}$ in the center mass frame.  The 
various polarization observables can also be expressed in terms
 of CGLN amplitudes, which can be found in Ref. \cite{tabakin}.

Therefore, it is more convenient to express the transition amplitudes
in the quark model in terms of the CGLN amplitudes, since the kinematics
in this framework is well known.  We start this procedure from the 
general quark-photon and quark meson interactions in the QCD Lagrangian
in Eq. \ref{1}.  By expanding the nonlinear field $\xi$ in Eq. \ref{2} 
in terms of the Goldstone boson fields $\pi$;
\begin{equation}\label{10}
\xi=1+i\pi/f+\dots,
\end{equation}
we obtain the standard pseudovector coupling at the tree level;
\begin{equation}\label{11}
H_m=\sum_j \frac 1{f_m} {\bar \psi}_j\gamma_\mu^j\gamma_5^j \psi_j
\partial^{\mu}\phi_m.
\end{equation}
The electromagnetic coupling is  
\begin{equation}\label{12}
H_e=-\sum_j e_j \gamma^j_\mu A^\mu ({\bf k}, {\bf r}).
\end{equation}

Because the baryon resonances in s- and u-channels are treated
 as three quark systems, the separation of the center of mass 
motion from the internal motions in the transition operators 
is crucial,  in particular, to reproduce the model independent 
low energy theorems\cite{zpli94} in the threshold 
pion-photoproduction and in the Compton 
scattering, $\gamma N\to \gamma N$\cite{zpli93}.  Thus, we take the
same approach as that in Refs. \cite{zpli93} and \cite{zpli94} to 
evaluate the contributions from resonances in s- and u-channels.  
Replacing the spinor $\bar \psi$ by $\psi^\dagger$ so 
that the $\gamma$ matrices are replaced by the matrix {\boldmath 
$\alpha$ \unboldmath}, the matrix elements for 
the electromagnetic interaction $H_{e}$ can be written as
\begin{eqnarray}\label{13}
\langle N_j|H_e |N_i\rangle & = &\langle N_j|\sum_j e_j \mbox {\boldmath 
$\alpha$ \unboldmath}_j \cdot \vep e^{i{\bf k}\cdot {\bf r}_j}  |N_i\rangle 
\nonumber \\
& = & i \langle N_j|[\hat H, \sum_j e_j {\bf r}_j \cdot 
\vep e^{i{\bf k}\cdot {\bf r}_j}] -\sum_j e_j {\bf r}_j\cdot \vep 
\mbox {\boldmath $\alpha$ \unboldmath}_j\cdot {\bf k}
e^{i{\bf k}\cdot {\bf r}_j}|N_i\rangle \nonumber \\
& = & i(E_j-E_i-\omega) \langle N_j| g_e|N_i\rangle +i\omega \langle N_j 
|h_{e}|N_i\rangle ,
\end{eqnarray}
where 
\begin{equation}\label{14}
\hat H= \sum_j (\mbox {\boldmath $\alpha$ \unboldmath}_j \cdot {\bf p}_j 
+\beta_jm_j)+\sum_{i,j}V({\bf r}_i-{\bf r_j})
\end{equation}
is the Hamiltonian for the composite system,
\begin{equation}\label{15}
g_e=\sum_j e_j {\bf r}_j \cdot \vep e^{i{\bf k}\cdot {\bf r}_j},
\end{equation}
\begin{equation}\label{16}
h_e=\sum_j e_j {\bf r}_j\cdot \vep (1-\mbox {\boldmath 
$\alpha$ \unboldmath}_j
\cdot \hat {\bf k})
e^{i{\bf k}\cdot {\bf r}_j},
\end{equation}
and $\hat {\bf k}=\frac {{\bf k}}{\omega_\gamma}$.
Similarly, we have
\begin{equation}\label{17}
\langle N_f|H_e |N_j\rangle 
 = i(E_f-E_j-\omega_\gamma) \langle N_f| g_e|N_j\rangle 
+i\omega_\gamma \langle N_f |h_{e}|N_j\rangle .
\end{equation}
Therefore the second and the third terms in Eq. \ref{4} can be written
as
\begin{eqnarray}\label{18}
{\cal M}_{23}^{\prime}=i \langle N_f|[g_e,H_m]|N_i\rangle 
+ i\omega_\gamma\sum_j\bigg \{ \frac {\langle N_f|H_m
|N_j\rangle \langle N_j |h_{e}|N_i\rangle }{E_i+\omega_\gamma-E_j} 
\nonumber \\  +\frac {\langle N_f|h_{e}|N_j
\rangle \langle N_j|H_m|N_i\rangle }{E_i-
\omega_m-E_j}\bigg \} \nonumber \\
= \langle N_f|{\cal M}_{seagul}^\prime|N_i\rangle + \langle N_f|{\cal M}_s 
|N_i\rangle+\langle N_f|{\cal M}_u|N_i\rangle
\end{eqnarray}
where the first term could also be regarded as the seagull term, 
and the ${\cal M}_s ({\cal M}_u)$ corresponds to the s(u)-channel
contributions. There are two very important consequences in this
 manipulation. First, the leading terms in the low energy 
theorem of the threshold pion photoproduction are present
 only in the leading Born terms,
which include the seagull term and the contributions from the nucleon
in the s- and u-channel, while the resonance contributions are only 
present at higher order\cite{zpli94}.  Second, the nonrelativistic 
expansion for $h_e$ in Eq. \ref{16} becomes\cite{zpli93,zpli94}
\begin{equation}\label{19}
h_e=\sum_j \left [ e_j {\bf r}_j \cdot \vep 
-\frac {e_j}{2m_j}{\vsig_j \cdot (\vep\times \hat
 {\bf k})} \right ]e^{i{\bf k}\cdot {\bf r}_j},
\end{equation}
which $h_e$ is only expanded to order $1/m_q$, and it has been 
shown\cite{zpli94} that the expansion to order $1/m_q$ is 
sufficient to 
reproduce the low energy theorem for the threshold 
pion-photoproductions\cite{cgln}.  The procedure from 
Eq. \ref{13} to Eq. \ref{19} is equivalent to the prescription 
in ref. \cite{close},  in which the effects
of the binding potential is included in the transition operator
so that the first term in Eq. \ref{19} differs from $\frac 1
{m_q}{\bf p}_j\cdot \vep$ used in Refs  \cite{cko} and \cite{simon}.   

The corresponding meson-coupling is 
\begin{eqnarray}\label{20}
H_{m}^{nr} =\sum_j \bigg \{ \frac {\omega_m}{E_f+M_f} \vsig_j\cdot 
{\bf P}_f+\frac {\omega_m}{E_i+M_i}\vsig_j \cdot {\bf P}_i-
\vsig_j\cdot {\bf q}\nonumber \\ +\frac {\omega_m}{2\mu_q}\vsig_j
\cdot {\bf p}_j \bigg \}\frac {\hat I_j}{g_A}e^{-i{\bf q}\cdot {\bf r}_j}
\end{eqnarray}
where $\omega_m$ is the energy of the emitting mesons
and ${\hat I_j}$ is an isospin operator.  The factor 
$\frac 1{\mu_q}$ in Eq. \ref{20} is a reduced mass at the quark level, 
which equals $\frac 1{\mu_q}=\frac 1{m_s}+\frac 1{m_q}$ for kaon 
productions and $\frac 1{\mu_q}=\frac 2{m_q}$ for $\eta$ and $\pi$
 productions.  The first three terms 
in Eq. \ref{20} correspond to the center of mass motion, and the last 
term represents the internal transition.   The constant $g_A$ is related 
to the axial vector coupling, and defined as
\begin{equation}\label{21}
\langle N_f| \sum_j {\hat I_j} {\vsig_j}
|N_i\rangle =g_A\langle N_f|{\vsig}|N_i\rangle .
\end{equation}
where  $\vsig$ is the total spin operator of the initial and final states
with spin 1/2.  The isospin operator ${\hat I_j}$ in Eq. \ref{20}
is 
\begin{equation}\label{22}
{\hat I_j}=\left \{ \begin{array}{r@{\quad}l} a^\dagger_j(s)a_j(u)
 & \mbox{ for 
$K^+$} \\ a^\dagger_j(s)a_j(d) & \mbox{ for $K^0$}  \\
a^\dagger_j(d) a_j (u) & \mbox{ for $\pi^+$}\\
-\frac 1{\sqrt{2}}(a^\dagger_j(u) a_j (u)-a^\dagger_j(d) a_j (d)) 
& \mbox{ for $\pi^0$}\\
1 & \mbox{ for $\eta$} 
\end{array}\right.  ,
\end{equation}
where $a^\dagger_j(s)$ and $a_j(u)$ or $a_j(d)$ are the creation and 
annihilation operator for the strange and up or down quarks, and
$I_j$ is simply an unit operator for the $\eta$ production so that
the factor $g_A$ is $1$.  The values of $g_A$ are listed in Table 1 
for each reaction in the $SU(6)$ symmetry limit.

\subsubsection*{\bf 3.1 The Seagull Term}

The amplitudes for the seagull term is 
\begin{equation}\label{24}
{\cal M}_s=-F({\bf k},{\bf q})e_m \left [ 1+\frac {\omega_m}2 \left 
(\fn+\fs\right )\right ]\vsig\cdot\vep.
\end{equation}
where $e_m$ is the charge of outgoing mesons, and the form 
factor is
\begin{equation}\label{25}
F({\bf k},{\bf q})=exp\left ( -\frac {({\bf k}-{\bf 
q})^2}{6 \alpha^2}\right )
\end{equation}
in the harmonic oscillator basis, where $\alpha$ is the oscillator 
strength.   

The seagull term in the chiral quark model is generated by the
gauge transformation of the QCD Lagrangian in Eq. \ref{1}.  
It produces the leading term\cite{zpli94} in the low energy 
theorem in the threshold pion photoproduction.  Therefore,
it plays a dominant role in the meson photoproductions of nucleons
in the low energy region.  The form factor in Eq. \ref{25} also makes
this term peaked at the forward angle for finite ${\bf k}$ and ${\bf q}$.
This leads to an interesting prediction for meson photoproductions in 
the chiral quark model; the differential cross sections for the
charged meson productions without contributions from
 isospin 3/2 resonances should be forward peaked above the threshold 
because of the dominance of seagull term in the low energy region.  
The data in the processes, $\gamma p\to K^+\Lambda$ and $\gamma p\to
\eta p$, are consistent with this conclusion,  in which the $K^+$ 
production is strongly forward peaked, while the $\eta$ production does
not exhibit the forward peaking at all.  This feature is quite unique 
in the chiral quark model,  it is a combination of the 
QCD Lagrangian and the integration of the spatial wavefunctions in 
the initial and final states,  which does not exist in the traditional 
effective Lagrangian approaches at hadronic level.

\subsubsection*{\bf 3.2 The U-channel resonance contribution}

We show the amplitudes for the u-channel $\Lambda$ and $\Sigma$ 
resonances in Kaon productions and for the u-channel nucleon in $\eta$ and
$\pi$ productions in Appendix.  The calculation of ${\cal M}_u$  
in Eq. \ref{18} for the excited states follows a procedure similar to 
that used in the Compton Scattering ($\gamma N\to \gamma N$)\cite{zpli93}.  
Replacing the outgoing photon operator $h_e$ in the Compton 
Scattering by $H_m^{nr}$ in Eq. \ref{20}, then the ${\cal M}_u$ is
\begin{equation}\label{23}
{\cal M}_u=\left ({\cal M}^3_u+{\cal M}^2_u\right )
e^{-\frac {{\bf k}^2+{\bf q}^2}{6\alpha^2}}
\end{equation}
where 
\begin{eqnarray}\label{231}
\frac {{\cal M}_u^3 g_A}3 = i
\frac {e_3 I_3}{2m_q} \vsig_3 \cdot (\vep \times {\bf k}) \vsig_3 \cdot
{\bf A} F^0(\frac {{\bf k}\cdot {\bf q}}{3
\alpha^2}, P_f\cdot k) \nonumber \\
-\frac {e_3I_3}6\left [ \frac {\omega_\gamma \omega_m}{\mu_q} \left (
 1+\frac {\omega_\gamma }{2m_q} \right ) \vsig_3 \cdot \vep + \frac 
{2\omega_\gamma}{\alpha^2} \vsig_3 \cdot {\bf A} \vep \cdot {\bf q}\right ] 
F^1(\frac {{\bf k}\cdot {\bf q}}{3
\alpha^2}, P_f\cdot k)
\nonumber \\
-\frac {\omega_\gamma \omega_m}{18\mu_q\alpha^2} e_3 I_3 \vsig_3
\cdot {\bf k} \vep \cdot {\bf q} F^2(\frac {{\bf k}\cdot {\bf q}}{3
\alpha^2}, P_f\cdot k),
\end{eqnarray} 
which corresponds to the outgoing meson and incoming photon absorbed
and emitted by the same quark, and 
\begin{eqnarray}\label{232}
\frac {{\cal M}_u^2 g_A}6 = i
\frac {e_2 I_3}{2m_q} \vsig_2 \cdot (\vep \times {\bf k}) \vsig_3 \cdot
{\bf A} F^0(-\frac {{\bf k}\cdot {\bf q}}{6
\alpha^2}, P_f\cdot k) \nonumber \\
+\frac {e_2I_3}{12}\left [ \frac {\omega_\gamma \omega_m}{\mu_q}\left (
 \vsig_3 \cdot \vep + \frac 1{2m_q}
\vsig_2 \cdot (\vep \times {\bf k})\vsig_3\cdot {\bf k} \right ) +\frac 
{2\omega_\gamma}{\alpha^2} \vsig_3 \cdot {\bf A} \vep 
\cdot {\bf q}\right ] \nonumber \\
\times F^1(-\frac {{\bf k}\cdot {\bf q}}{6
\alpha^2}, P_f\cdot k)
-\frac {\omega_\gamma \omega_m}{72\mu_q\alpha^2} e_2 I_3 \vsig_3
\cdot {\bf k} \vep \cdot {\bf q} F^2(-\frac {{\bf k}\cdot {\bf q}}{6
\alpha^2}, P_f\cdot k),
\end{eqnarray} 
which corresponds to the incoming photon and outgoing meson absorbed
and emitted by different quarks.
The vector ${\bf A}$ in Eqs. \ref{231} and \ref{232} are defined as
\begin{equation}\label{28}
{\bf A}=-\omega_m\left (\fn +\fs\right ){\bf k}-\left ( \omega_m\fs
+1\right ){\bf q}.
\end{equation}
Notice that the initial nucleon and the intermediate states have the 
c.m. momenta $-{\bf k}$ and $-{\bf k}-{\bf q}$ respectively. 
The function $F^l(x, y)$ in Eqs. \ref{32} and \ref{33} is the product 
of the spatial integral and the propagator for the excited states, 
it can be written as
\begin{equation}\label{34}
F^l(x, y)=\sum_{n\ge l} \frac {M_{n}}{ (n-l)! (y+n\delta M^2)} x^{n-l}, 
\end{equation}
where $n\delta M^2=(M_n^2-M^2_f)/2$ represents the mass
difference between the ground state and excited states with the total 
excitation quantum number $n$ in the harmonic oscillator basis. 

The ${\cal M}_u^3$ in Eq. \ref{231} and ${\cal M}_u^2$ in Eq. \ref{232}
are the u-channel operators at the quark level.  They are the masteer 
equations for all pseudo scalar meson photoproduction.  To derive the
amplitudes for a particular reaction, one has to transform Eqs. \ref{231}
and \ref{232} into the more familiar CGLN amplitudes at the hadron level.
We have
 \begin{eqnarray}\label{32}
\frac {{\cal M}^3_u}{g^u_3}=\frac {1}{2m}
\left [i g_v {\bf A}\cdot (\vep\times {\bf k})+\vsig\cdot ({\bf 
A}\times (\vep\times {\bf k}))\right ]F^0(\frac {{\bf k}\cdot {\bf q}}{3
\alpha^2}, P_f\cdot k) \nonumber \\
-\frac 1{6}\left [\frac {\omega_m\omega_{\gamma}}{\mu_q}\left (1+\frac 
{\omega_{\gamma}}{2m_q}\right )\vsig \cdot \vep+\frac 
{2\omega_\gamma}{\alpha^2}\vsig\cdot {\bf A}\vep\cdot {\bf q}\right ]
  F^1(\frac {{\bf k}\cdot {\bf q}}{3\alpha^2}, P_f\cdot k)
\nonumber \\ -\frac {\omega_m\omega_\gamma}{18\alpha^2\mu_q}\vsig\cdot {\bf 
k}\vep\cdot {\bf q}
  F^2(\frac {{\bf k}\cdot {\bf q}}{3
\alpha^2},P_f\cdot k),
\end{eqnarray}
and 
\begin{eqnarray}\label{33}
\frac {{\cal M}^2_u}{g_2^u}
= -\frac 1{2m_q} \left [ -ig_v^\prime {\bf A}\cdot (\vep \times 
{\bf k})+g_a^\prime \vsig \cdot ({\bf A}\times (\vep\times {\bf k}))\right ]
 F^0(\frac {-{\bf k}\cdot {\bf q}}{6\alpha^2}, P_f\cdot k)\nonumber 
\\
+\frac 1{12}\left [\frac {\omega_m\omega_{\gamma}}{\mu_q}
\left (1+g_a^\prime\frac {\omega_{\gamma}}{2m_q}\right )\vsig\cdot\vep+\frac 
{2\omega_{\gamma}}{\alpha^2} \vsig \cdot {\bf A}\vep\cdot {\bf q}\right ]
F^1(-\frac {{\bf k}\cdot {\bf q}}{6
\alpha^2},P_f\cdot k) \nonumber  \\ -
\frac {\omega_m\omega_\gamma}{72\alpha^2 \mu_q}\vsig \cdot {\bf 
k}\vep\cdot {\bf q} F^2(-\frac {{\bf k}\cdot {\bf q}}{6\alpha^2},P_f\cdot k).
\end{eqnarray}
The various g-factors in Eqs. \ref{32} and \ref{33} are defined as
\begin{equation}\label{36}
g^u_3 =\langle N_f| \sum_j e_j{\hat I}_j \sigma_j^z | N_i\rangle /g_A,
\end{equation}
\begin{equation}\label{37}
g^u_2=\langle N_f| \sum_{i\not = j} e_j{\hat I}_i \sigma_i^z | N_i
\rangle /g_A,
\end{equation}
and
\begin{equation}\label{35}
 g_v=\langle N_f| \sum_j e_j {\hat I}_j  | N_i\rangle / g^u_3 g_A
 ,
\end{equation}
where $g_A$ is given in Eq. \ref{21}.  The factors $g_v^\prime$ and 
$g_a^\prime$ in Eq. \ref{33} come from 
\begin{equation}\label{38}
\frac 1{g^u_2g_A}
\langle N_f| \sum_{i\not = j} e_j{\hat I}_i \vsig_i \cdot {\bf A}
 \vsig_j \cdot {\bf B} | N_i\rangle 
=\langle N_f| g_v^\prime {\bf A}\cdot {\bf B} + i g_a^\prime
  \vsig \cdot ({\bf A}\times
{\bf B}) | N_i\rangle ,
\end{equation}
where ${\bf A}$ and ${\bf B}$ are vectors, and $\vsig$ is the total 
spin operator for spin 1/2 baryons.  Thus, we have 
\begin{equation}\label{381}
g_v^\prime = \frac 1{3g_2^ug_A} \langle N_f|\sum_{i\not = j} e_j{\hat I}_i
\vsig_i \cdot \vsig_j | N_i\rangle ,
\end{equation}
and 
\begin{equation}\label{382}
g_a^\prime = \frac 1{2g_2^ug_A} \langle N_f|\sum_{i\not = j} e_j{\hat I}_i
(\vsig_i \times \vsig_j )_z| N_i\rangle .
\end{equation}
The numerical values of these g-factors in Eqs. \ref{32} and \ref{33} 
depend on the detialed structures of final state wavefunctions, and 
they are presented in Table 1 in the $SU(6)$ symmetry limit.

The first term in Eqs. \ref{32} and \ref{33} corresponds to the 
correlation between the magnetic transition and the c.m. motion of 
the kaon transition operator, it contributes to the leading 
Born term in the U-channel. The second term
is the correlations among the internal and c.m. motions of 
the photon and meson transition operators, they only contribute to the 
transitions between the ground and $n\ge 1$ excited states in the 
harmonic oscillator basis.  
The last term in both equations represents
the correlation of the internal motions between the photon and meson 
transition operators, which only contribute to the 
transition between the ground and $n\ge 2$ excited states.  
An interesting observation from these expressions is that  the transition 
matrix elements ${\cal M}_u^3$ and ${\cal M}^2_u$ correspond to the
incoming photons and outgoing kaons being absorbed and emitted by the same 
and different quarks,  and they differ by a factor $\left (-\frac 12\right 
)^n$.  Thus the transition matrix element ${\cal M}^3_u$ becomes 
dominant as the quantum number $n$ increases. 

Eqs. \ref{32} and \ref{33} can be summed up to any quantum number $n$,
however, the excited states with large quantum number $n$ become
less significant for the u-channel resonance contributions.  Thus, we only 
include the excited states with $n\le 2$, which is the minimum number 
required for the contributions from every term in Eqs. \ref{32} and 
\ref{33}.  Physically, this corresponds to the average 
sum of the contributions from every resonance with the total excitation 
number $n=1$ and $2$. 
The orbital excited $n\ge 1$ resonances are treated as degenerate, 
since their contributions in the u-channel are much less sensitive
to the detail structure of their masses than those in the s-channel.
However, the contributions from $\Sigma^*(1385)$ and $\Delta (1232)$
to the $K$ and $\pi$ photoproductions should be separated from the 
$\Lambda$, $\Sigma$ and nucleon for $n=0$, as their masses differs
significantly.  The amplitude ${\cal M}_u$ for $n=0$ is 
\begin{eqnarray}\label{384}
{\cal M}_u^{n=0}=\frac 1{2m_q}\frac {e^{-\frac {{\bf q}^2+{\bf 
k}^2}{6\alpha^2}}}{P_f\cdot k+\delta M^2/2} [ 
i(g^u_3g_v+g^u_2g_v^\prime) {\bf A}\cdot (\vep \times 
{\bf k}) \nonumber \\
+ (g^u_3-g^u_2 g_a^\prime) \vsig \cdot ({\bf A}\times 
(\vep\times {\bf k}))  ]
\end{eqnarray}
Eq. \ref{384} represents the sum of the
contributions from both spin 1/2 and 3/2 states in ${\bf 56}$ multiplet, 
such as the $\Lambda $, $\Sigma $ and $\Sigma^*$ states in kaon 
photoproduction.  Thus, the contributions from spin 3/2 states 
such as the $\Sigma^*$ and $\Delta$ can be obtained by 
subtracting the contributions of spin $1/2$  intermediate 
states from the total $n=0$ amplitudes.  The amplitude for spin 1/2
intermediate state is
\begin{equation}\label{383}
\langle N_f | h_e | N(J=1/2) \rangle \langle N(J=1/2) 
| H_m | N_i \rangle
= - \langle N_f |\mu  \vsig \cdot (\vep \times {\bf k})
\vsig \cdot {\bf A}| N_i \rangle ,
\end{equation} 
in which we only considered the contributions of the magnetic term
in $h_e$.  The magnetic moment $\mu$ in Eq. \ref{383} is 
\begin{equation}\label{394}
\mu =\left \{ \begin{array}{r@{\quad}l}
\mu_{\Lambda} + \frac {g_{K\Sigma N}}{g_{K\Lambda N}}
 \mu_{\Lambda \Sigma} & \mbox{ for $\gamma N\to K\Lambda$} \\
\mu_{\Sigma^0} + \frac {g_{K\Lambda N}}{g_{K\Sigma N}}
\mu_{\Lambda \Sigma} &  \mbox{ for $\gamma N\to K \Sigma^0$} \\
\mu_{N_f} & \mbox{ for other $\gamma N\to m_{0^{-+}} N_f$}
\end{array} \right. .
\end{equation}
We have 
\begin{eqnarray}\label{385}
{\cal M}_u^{J=3/2}=\frac 1{2m_q}\frac {e^{-\frac {{\bf q}^2+{\bf 
k}^2}{6\alpha^2}}}{P_f\cdot k+\delta M^2/2} [ 
i(g^u_3g_v+g^u_2g_v^\prime) {\bf A}\cdot (\vep \times 
{\bf k}) \nonumber \\
+ (g^u_3-g^u_2 g_a^\prime) \vsig \cdot ({\bf A}\times 
(\vep\times {\bf k})) -i\mu  \vsig \cdot (\vep \times {\bf k})
\vsig \cdot {\bf A} ]
\end{eqnarray} 
We find that the general form  of the CGLN amplitudes for the 
$J=3/2$ states with $n=0$  is 
\begin{equation}\label{29}
{\cal M}_u^{J=3/2}=\frac {M_fg_Se^{-\frac {{\bf q}^2+{\bf 
k}^2}{6\alpha^2}}}{M_N(P_f\cdot k+\delta M^2_{J=3/2}/2)}
\left [i2\vsig \cdot {\bf A} \vsig\cdot 
(\vep\times {\bf k})+\vsig\cdot ({\bf A}\times (\vep\times {\bf k}))\right ]
\end{equation}
where $\delta  M^2_{J=3/2}=M^2_{J=3/2}-M^2_f$. The factor $g_S$  
is also listed in Table 1 for each reaction.  The structure in Eq. \ref{29}
is different from that of the CGLN amplitude for the spin 3/2 resonance
in the s-channel; there are both $M^-_1$ and $M^+_1$ components 
in Eq. \ref{29}, while the spin 3/2 resonance in the s-channel
only has  $M^+_1$ transition\cite{tabakin}.
  
\subsubsection*{\bf 3.3 The S-channel resonance contribution}

The amplitude of the nucleon pole term is presented in the Appendix.
The S-channel resonance contributions comes from the second term in 
Eq. \ref{18}, the derivation of this term follows the analytic 
procedure for the Compton scattering in  Ref. \cite{zpli93}. 
 Replacing the outgoing photon vertex in Compton scattering in Ref. 
\cite{zpli93} by the meson transition operator in Eq. \ref{20}, we 
find that the general expression for the excited resonances in the 
 s-channel can be written as
\begin{equation}\label{39}
{\cal M}_R=\frac {2M_R}
{s-M_R(M_R-i\Gamma({\bf q}))}e^{-\frac {{\bf k}^2+{\bf q}^2}
{6\alpha^2}}{\cal O}_R,
\end{equation}
where $\sqrt {s}=E_i+\omega_{\gamma}=E_f+\omega_{m}$ 
is the total energy of the system, and ${\cal O}_R$ is determined 
by the structure of each resonance.  Eq. \ref{39} shows that there should be 
a form factor,  $e^{-\frac {{\bf k}^2+{\bf
q}^2}{6\alpha^2}}$ in the harmonic oscillator basis, even in the real 
photon limit.    
$\Gamma({\bf q})$ in Eq. \ref{39} is the total width of the resonance, 
and a function of the final state momentum ${\bf q}$.  For a resonance 
decaying into a two-body final state with relative angular momentum $l$,
the decay width $\Gamma({\bf q})$ is
\begin{equation}\label{40}
\Gamma({\bf q})= \Gamma_R \frac {\sqrt {s}}{M_R} \sum_{i} x_i 
\left (\frac {|{\bf q}_i|}{|{\bf q}^R_i|}\right )^{2l+1} 
\frac {D_l({\bf q}_i)}{D_l({\bf q}^R_i)},
\end{equation}
with 
\begin{equation}\label{41}
|{\bf q}^R_i|=\sqrt{\frac 
{(M_R^2-M_N^2+M_i^2)^2}{4M_R^2}-M_i^2},
\end{equation}
and 
\begin{equation}\label{42}
|{\bf q}_i|=\sqrt{\frac 
{(s-M_N^2+M_i^2)^2}{4s}-M_i^2},
\end{equation}
where $x_i$ is the branching ratio of the resonance decaying into a 
meson with mass $M_i$ and a nucleon, and $\Gamma_R$ is the total decay width 
of the S-channel resonance with the mass $M_R$.  The 
function $D_l({\bf q})$ in Eq. \ref{40} 
is called fission barrier\cite{bw}, and wavefunction 
dependent. For the meson transition operator in Eq.\ref{20}, the 
$D_l({\bf q})$ in the harmonic oscillator basis has
 the form\cite{simon}
\begin{equation}\label{43}
D_l({\bf q})=exp\left (-\frac {{\bf q}^2}{3\alpha^2}\right ),
\end{equation}
which is independent of $l$. A similar formula used in I=1
 $\pi\pi$ and p-wave $I=1/2$ $K\pi$ scattering was found 
in excellent  agreement with data in the $\rho$ and $K^*$ 
meson region\cite{barnes}.  In principle, the branching ratio
$x_i$ should be evaluated in the quark model.  However, there are
very large uncertainties in most quark model evaluation as the 
coupling constant, such as $\alpha_{\eta NN}$ and 
$\alpha_{K\Lambda N}$, are not well determined.  Our results 
in the $\eta$ and kaon photproductions could provide a guide for 
the future investigations, which in turn will determine the branching
ratios $x_i$ more precisely.  Therefore, we simply set 
$x_{\pi}=x_{\eta}=0.5$ for the resonance $S_{11}(1535)$, while 
$x_{\pi}=1.0$ for the rest of the resonances as a first order 
approximation, as the resonance decays are dominated by the
 pion channels except the resonance $S_{11}(1535)$ whose branching
ratio in $\eta N$ channel is around 50 percent.  The results in $\eta$
and kaon photoproductions suggest that they are not sensitive to the
quantity $x_i$, as we know qualitatively that $x_i$ is small in the
$\eta N$ and $K Y$ channels except the case of $S_{11}(1535)$.

Our investigation in the $\eta$ photoproduction\cite{zpli952} has shown
that the momentum dependence of the decay width for the s-channel 
resonances is  very important in extracting the properties of the resonance
$S_{11}(1535)$ from the data in the threshold $\eta$ production,  
it is also an important procedure
to ensure the unitarity of the total transition amplitudes\cite{muko} 
approximately.  This has not been taken into account in many calculations of
the kaon and $\eta$ production within the effective Lagrangian approach,
which leads to a larger theoretical uncertainty that has not been fully 
investigated.

At the quark level, the operator ${\cal O}_R$ for a given $n$ in 
the harmonic oscillator basis is
\begin{equation}\label{50}
{\cal O}_{n}={\cal O}_n^2 +{\cal O}_n^3
\end{equation}
where the amplitudes ${\cal O}_n^2$ and ${\cal O}_n^3$ have the same 
meaning as the amplitudes ${\cal M}_u^2$ and ${\cal M}_u^3$ in Eqs. 
\ref{231} and \ref{232}.  Following the same procedure used in the Compton
Scattering\cite{zpli93}, we have
\begin{eqnarray}\label{241}
\frac {{\cal O}_n^3 g_A}3 = -i\frac { I_3e_3}{2m_q} \vsig_3 \cdot
{\bf A}\vsig_3 \cdot (\vep \times {\bf k}) \frac 1{n!}\left (\frac 
{{\bf k}\cdot {\bf q}}{3\alpha^2}\right )^n \nonumber \\
+\frac {e_3I_3}6\left [ \frac {\omega_\gamma \omega_m}{\mu_q} \left (
 1+\frac {\omega_\gamma }{2m_q} \right ) \vsig_3 \cdot \vep + \frac 
{2\omega_\gamma}{\alpha^2} \vsig_3 \cdot {\bf A} \vep \cdot {\bf q}\right ] 
\frac 1{(n-1)!}\left (\frac {{\bf k}\cdot {\bf q}}{3\alpha^2}\right )^{n-1}
\nonumber \\
+\frac {\omega_\gamma \omega_m}{18\mu_q\alpha^2} e_3 I_3 \vsig_3
\cdot {\bf k} \vep \cdot {\bf q}\frac 1{(n-2)!}
  \left (\frac {{\bf k}\cdot {\bf q}}{3\alpha^2}\right )^{n-2} ,
\end{eqnarray} 
 and 
\begin{eqnarray}\label{242}
\frac {{\cal O}_n^2 g_A}6 = -i
\frac {e_2 I_3}{2m_q} \vsig_2 \cdot (\vep \times {\bf k}) \vsig_3 \cdot
{\bf A} \frac 1{n!}\left (\frac {-{\bf k}\cdot {\bf q}}{6\alpha^2}\right )^n
 \nonumber \\
-\frac {e_2I_3}{12}\left [ \frac {\omega_\gamma \omega_m}{\mu_q}\left (
 \vsig_3 \cdot \vep + \frac 1{2m_q}
\vsig_2 \cdot (\vep \times {\bf k})\vsig_3\cdot {\bf k} \right ) +\frac 
{2\omega_\gamma}{\alpha^2} \vsig_3 \cdot {\bf A} \vep 
\cdot {\bf q}\right ] \nonumber \\
\times \frac 1{(n-1)!}\left (\frac {-{\bf k}\cdot {\bf q}}{6
\alpha^2}\right )^{n-1} 
+\frac {\omega_\gamma \omega_m}{72\mu_q\alpha^2} e_2 I_3 \vsig_3
\cdot {\bf k} \vep \cdot {\bf q} \frac 1{(n-2)!}\left (\frac {-{\bf k}\cdot 
{\bf q}}{6\alpha^2}\right )^{n-2},
\end{eqnarray} 
where the vector ${\bf A}$ for the s-channel is 
\begin{equation}\label{54}
{\bf A}=-\left (\frac {\omega_m}{E_f+M_f}+1\right ){\bf q}.
\end{equation}
One can transform Eqs. \ref{241} and \ref{242} into more familiar
CGLN amplitudes, and we find
\begin{eqnarray}\label{51}
\frac {{\cal O}^3_n}{g^s_3}
= -\frac {1}{2m_q}
\left [ig_v {\bf A}\cdot (\vep\times {\bf k})-\vsig\cdot ({\bf 
A}\times (\vep\times {\bf k}))\right ]\frac 1{n!}\left (\frac 
{{\bf k}\cdot {\bf q}}{3\alpha^2}\right )^n \nonumber \\
+\frac 1{6}\left [\frac {\omega_m\omega_{\gamma}}{\mu_q}\left (1+\frac 
{\omega_{\gamma}}{2m_q}\right )\vsig \cdot \vep+\frac 
{2\omega_{\gamma}}{\alpha^2}\vsig\cdot {\bf A}\vep\cdot 
{\bf q}\right ]\frac 1{(n-1)!}
  \left (\frac {{\bf k}\cdot {\bf q}}{3\alpha^2}\right )^{n-1}
\nonumber \\ +\frac {\omega_m\omega_\gamma}{18\alpha^2\mu}\vsig\cdot {\bf 
k}\vep\cdot {\bf q}
\frac 1{(n-2)!} \left (\frac {{\bf k}\cdot {\bf q}}{3
\alpha^2}\right )^{n-2}
\end{eqnarray}
and
\begin{eqnarray}\label{52}
\frac {{\cal O}^2_n}{g^u_2}
= \frac 1{2m_q} \left [ -ig_v^\prime {\bf A}\cdot (\vep \times 
{\bf k})+g_a^\prime \vsig \cdot ({\bf A}\times (\vep\times {\bf k})\right ]
 \frac 1{n!}\left (\frac {-{\bf k}\cdot {\bf q}}{6\alpha^2}\right )^n
\nonumber \\
-\frac 1{12}\left [\frac {\omega_m\omega_{\gamma}}{\mu_q}
\left (1+g_a^\prime\frac {\omega_{\gamma}}{2m_q}\right )\vsig\cdot \vep+\frac 
{2\omega_{\gamma}}{\alpha^2} \vsig \cdot {\bf A}\vep\cdot {\bf q}\right ]
\frac 1{(n-1)!}\left (\frac {-{\bf k}\cdot {\bf q}}{6
\alpha^2}\right )^{n-1} \nonumber  \\
+\frac {\omega_m\omega_\gamma}{72\alpha^2 \mu}\vsig \cdot {\bf 
k}\vep\cdot {\bf q} \frac 1{(n-2)!}\left (\frac {-{\bf k}\cdot 
{\bf q}}{6\alpha^2}\right )^{n-2}.
\end{eqnarray}
where the g-factors in Eqs. \ref{51} and \ref{52} are defined in Eqs. 
\ref{36}-\ref{382} and given in Table 1, and 
\begin{equation}\label{53}
g^s_3=\langle N_f| \sum_j e_j{\hat I}_j \sigma_j^z | N_i\rangle /g_A
={e_m}+g^u_3,
\end{equation}
where $e_m$ is the charge of the outgoing mesons.
Thus,  the operator ${\cal O}_R$ in Eq. \ref{39}  has a general structure,
\begin{equation}\label{44}
{\cal O}_R=g_RA\left [f_1^R \vsig\cdot \vep
+ if_2^R {(\vsig \cdot {\bf q})\vsig \cdot ({\bf k}\times \vep)}
+f_3^R{\vsig \cdot {\bf k}}\vep\cdot {\bf q}+f_4^R{\vsig \cdot {\bf q}}
\vep\cdot {\bf q}\right ],
\end{equation}
for the pseudoscalar meson photoproductions, where $g_R$
 is an isospin factor, $A$ 
the meson decay amplitude, and $f_i^R$ ($i=1\dots 4$) is the photon 
transition amplitude.  The factor $g_R$ and the meson decay amplitude
$A$ in Eq. \ref{44} are determined by the matrix elements $\langle N_f |H_m
|N_j\rangle $ in Eq. \ref{4}; the factor $g_R$ represents the transition
in the spin-flavor space, and the amplitude $A$ is the integral of the
spatial wavefunctions.  

We shall discuss briefly how the CGLN amplitudes for each resonance
with $n=1$ could be extracted from Eq. \ref{51} and \ref{52}, as the
CGLN amplitudes for $J=3/2$ resonances with n=0 follow the same procedure
as that in the u-channel.  Since the amplitude ${\cal O}_{n=1}$ represents 
the sum of all resonances with $n=1$, we start with the reaction $\gamma p
\to K^+\Lambda$, in which the isospin 3/2 does not contribute.  Moreover,
the contributions from the states with quantum number $N(^4P_M)$ vanish
 as well
due to Moorhouse selection rule\cite{moor}
 for the electromagnetic transition $h_e$
in Eq. \ref{19}.  Thus, only the resonances with $N(^2P_M)$ 
contribute to the reaction $\gamma p\to K^+ \Lambda$.  
Substitute the g-factors for the reaction $\gamma p\to K^+ \Lambda$ into 
Eqs. \ref{51} and \ref{52}, we have 
\begin{eqnarray}\label{511}
{\cal O}_{n=1}=-\frac {1}{12m_q}
\left [i{\bf A}\cdot (\vep\times {\bf k})-\vsig\cdot ({\bf 
A}\times (\vep\times {\bf k}))\right ]\frac {{\bf k}\cdot {\bf q}}{\alpha^2}
\nonumber \\
+\frac 1{12}\left [\frac {\omega_m\omega_{\gamma}}{\mu}\left (1+\frac 
{\omega_{\gamma}}{2m_q}\right )\vsig \cdot \vep+\frac 
{2\omega_{\gamma}}{\alpha^2}\vsig\cdot {\bf A}\vep\cdot 
{\bf q}\right ],
\end{eqnarray}
in which only $S$ and $D$ partial waves are present.  Rewrite the quantity
${\cal O}_{n=1}$ into $S$ and $D$ waves, we find
\begin{equation}\label{512}
{\cal O}_{n=1}(S\ wave)=\frac {\omega_\gamma}{12}\left 
(1+\frac {\omega_\gamma}{2m_q}\right )\left (\frac {\omega_m}{\mu_q}+\frac 
2{3}\frac {{\bf A}\cdot {\bf q}}{\alpha^2}\right ) \vsig \cdot \vep ,
\end{equation}
and 
\begin{equation}\label{513}
{\cal O}_{n=1}(D\ wave) =\frac {-i}{12m_q} \vsig \cdot {\bf A} \vsig\cdot (\vep\times {\bf k}) -\frac {\omega_\gamma}{18}\frac {{\bf A}\cdot {\bf
 q}}{\alpha^2}\left 
(1+\frac {\omega_\gamma}{2m_q}\right )\vsig \cdot \vep + \frac 16 \frac 
{\omega_{\gamma}}{\alpha^2}\vsig\cdot {\bf A}\vep\cdot 
{\bf q} .
\end{equation}
Thus, the ${\cal O}_{n=1}(S\ wave)$ in Eq. \ref{512} represents the CGLN 
amplitude for the resonance $S_{11}$ with quantum number 
$N(^2P_M){\frac 12}^-$, while ${\cal O}_{n=1}(D\ wave)$ is the CGLN amplitude
for the resonance $D_{13}$ with quantum number $N(^2P_M){\frac 32}^-$, as
only the $S_{11}$ resonance with $N(^2P_M){\frac 12}^-$ and the $D_{13}$
 resonance with $N(^2P_M){\frac 32}^-$ contribute to 
$\gamma p \to K^+ \Lambda$.
The quantity $\left (\frac {\omega_m}{\mu_q}+\frac 2{3}\frac {{\bf A}\cdot
 {\bf q}}{\alpha^2}\right )$ in Eq. \ref{512} corresponds to the meson 
transition amplitude $A$ in Eq. \ref{44},  while the meson transition
amplitude $A$ for the $D-wave$ resonance is $\frac {|{\bf A}|}{|{\bf q}|}$. 
The amplitudes $A$ for the $S$ and $D$ waves have the same  expressions as
those in Table 1 of Ref. \cite{simon} with $g-\frac 13h=\frac 
{|{\bf A}|}{|{\bf q}|}$, and $h=\frac {\omega_m}{2\mu_q}$.  Note that 
${\bf A}$ has a negative sign,  this is consistent with the fitted values 
for $g-\frac 13h$ and $h$ in Ref. \cite{simon}.   The quantity
 $\frac {\omega_\gamma}{12}\left (1+\frac {\omega_\gamma}{2m_q}\right )$
in Eq. \ref{512} represents the photon transition amplitude $f_i^R$. It is
proportional to the helicity amplitude $A_{\frac 12}^p$ for the state 
$N(^2P_M){\frac 12}^-$ for the $h_e$ in Eq. \ref{19}\cite{zpli90}, as the 
CGLN amplitude for S-wave resonances is simply a product of photon and meson
transition amplitudes.  

Similarly, we find that the CGLN amplitude for the
$S_{31}$ resonance in the reaction $\gamma p\to K^+ \Sigma^0$ is
\begin{equation}\label{514}
{\cal O}_{n=1}(S\ wave)=\frac {\omega_\gamma}{6}\left 
(1-\frac {\omega_\gamma}{6m_q}\right )\left (\frac {\omega_m}{\mu_q}+\frac 
2{3}\frac {{\bf A}\cdot {\bf q}}{\alpha^2}\right ) \vsig \cdot \vep .
\end{equation}
This shows that the amplitude $A$ is the same for both reactions, $\gamma 
p\to K^+\Lambda$ and $\gamma p\to K^+ \Sigma^0$.  In fact, it has been shown
in Ref. \cite{simon} that the meson transition amplitude $A$ is independent 
of not only a particular reaction but also $SU(6)$ symmetry so that the
 resonances belonging to $(56)$ and $(70)$ multiplets with the same angular 
momentum $L=2$ are governed by the same meson decay amplitude $A$.  In other 
words  the amplitude $A$ in Eq. \ref{44} is universal for the pseudoscalar 
meson decay processes.
Thus,  we present the amplitude $A$ in the simple harmonic oscillator basis 
in Table 2, in  which the amplitude $A$ depends on the total excitation 
$n$ and the orbital angular momentum $L$.   The  relative angular momentum 
of the final decay products is expressed in terms of 
the partial wave language in Table 2, which the $S$, $P$, $D$ and $F$ waves
denote the relative angular momentum 0, 1, 2 and 3 between the 
final decay products.  

Thus, the advantages of Eq. \ref{44} are that only the factor $g_R$ is 
determined by a particular reaction, while the amplitude $A$ is universal, and
the photon transition amplitudes $f^R_i$ only depend on the initial proton
and neutron targets.   We show the photon transition amplitudes $f_i^R$ for 
each resonance with $n\le 2$ in Table 3 for the proton target and 
Table 4 for the neutron target.
They are usually expressed in terms of helicity amplitudes, $A_{1/2}$ 
and $A_{3/2}$, and the connection between the two representations can be 
established,  which has been discussed extensively in Refs. \cite{zpli951}
and \cite{zpli952} for proton targets.  Here we discuss some important
features of the CGLN amplitudes for neutron targets and their relation
to those for proton targets. 
 A very important example is the contributions from the resonances 
belonging to $(70, ^4N)$ representation for neutron targets, of which
the transition amplitudes for proton targets are zero
due to the Moorhouse selection rule\cite{moor} if one uses the 
nonrelativistic transition operator in Eq. \ref{20}.
There are three important negative parity baryons that belong to 
the $(70, ^4N)$ multiplet in the naive quark model,  which 
correspond to $S_{11}(1650)$, $D_{13}(1700)$ and $D_{15}(1675)$. 
In particular, the contributions from the resonance $D_{15}(1675)$
 are quite large.   In general, the 
constraints on the phototransition in terms of the helicity amplitudes, 
$A_{\frac 12}$ and $A_{\frac 32}$ in the $SU(6)\otimes O(3)$ symmetry 
limit can also be applied to the corresponding CGLN amplitudes.  For example, 
the helicity amplitude $A_{\frac 32}$ for the resonance $D_{13}(1520)$ 
 classified as $N(^2P_M){\frac 32}^-$ in the quark model has a 
simple relation\cite{close,simon}; 
\begin{equation}\label{466}
A^p_{\frac 32}(D_{13}(1520))=-A^n_{\frac 32}(D_{13}(1520)),
\end{equation}
and we find the same relation for the corresponding CGLN amplitudes 
$f_4$ 
\begin{equation}\label{47}
f_4^p(D_{13}(1520))=-f_4^n(D_{13}(1520))
\end{equation}
between protons and neutrons.  

The CGLN amplitudes for three S-wave resonances show a more explicit 
connection with the corresponding helicity amplitudes.   Because 
only $f_1^R$ is present for the S-wave resonances, it is proportional 
to the helicity amplitude $A_{1/2}$. Thus, we have 
\begin{equation}\label{488}
\frac {f_1^R(\gamma p\to S_{11})}{f_1^R(\gamma n\to S_{11})}
=\frac {A^p_{1/2}(S_{11})}{A^n_{1/2}(S_{11})},
\end{equation}
and the comparison between the CGLN amplitudes in Table 4 and 5 
and the corresponding helicity amplitudes in Ref. \cite{close} 
show that this is indeed the case.

For the excited positive parity baryon resonances, the helicity 
amplitude $A^n_{\frac 32}$ vanishes for the states $N(^2D_s)$, and 
corresponding CGLN amplitude $f^n_4$ is zero for these states as well.  The 
ratio of the helicity amplitudes $A_{\frac 12}$ between the proton and the
neutron targets for the resonance $P_{11}(1440)$ is the same as 
the ratio of the 
CGLN amplitude $f_2$, which corresponds to the $M_1^-$ transition according
to the multipole decomposition of the CGLN amplitude\cite{tabakin}.
There are also contributions from the states $N(^4D_M)$, which are in the 
same $SU(6)$ representation as the states $N(^4P_M)$ so that the Moorhouse 
selection rule is also true for these states.   However, we  find that 
only the CGLN amplitudes for the state $F_{17}(1990)$ is relatively strong, 
and there is little evidence for other resonances below 2 GeV.   Therefore, 
only the contribution from $F_{17}(1990)$ will be taken into account in our 
calculation.  The CGLN amplitudes for resonances 
$P_{33}$ belonging to the $56$ representations satisfy the relation
\begin{equation}\label{48}
\frac {f_1^R}{{\bf q\cdot k}}=-f_3^R=\frac 32f_2^R.
\end{equation}
According to the multipole decomposition of the CGLN 
amplitudes\cite{tabakin}, Eq. 
\ref{47} corresponds to the $M^+_1$ transition which also leads to the 
relation\cite{close,simon}
\begin{equation}\label{49}
A_{1/2}=\frac 1{\sqrt {3}}A_{3/2}
\end{equation}
between the two helicity amplitudes.  This is certainly true for
the resonances $P_{33}(1232)$ and $P_{33}(1600)$ in the symmetry 
limit.

We present the quark model results 
of the $g_R$-factor for the pseudoscalar meson  photoproductions
in Table 5.   It represents the relative strength and phases of 
the contributions from different resonances comparing to the contributions 
from the nucleon which belongs to the $(56,N)$ multiplet in $SU(6)$ symmetry. 
Notice that for a given $SU(3)$ representation, 
the factor $g_R$ is determined
by the C-G coefficient in the isospin coupling  between the meson and the 
final baryon state, 
\begin{equation}\label{45}
g_R \propto \langle I_m, I_m^z, I_f, I_f^z | I_R, 
I_m^z+I_f^z\rangle /g_A,
\end{equation}
where $I_m^z$ and $I_m$ are the isospin for the outgoing mesons, 
$I_f$ and $I_f^z$ are the isospin quantum numbers for final baryons, 
and $I_R$ is the isospin of s-channel resonances.  Thus, we have the relation 
\begin{equation}\label{46}
\frac { g_R(\gamma p \to K^0\Sigma^+)}{g_R(\gamma n\to K^+\Sigma^-)}=
\frac {\langle \frac 12, -\frac 12, 1, 1 | I_R, \frac 12\rangle g_A(\gamma n
\to K^+\Sigma^-)}{\langle \frac 12, \frac 12, 1, -1 | I_R, -\frac 12\rangle
g_A(\gamma p\to K^0\Sigma^+)}=(-1)^{I_R-\frac 12},
\end{equation}
in which the additional minus sign comes from the ratio of the $g_A$.  
Results in Table 2 show that the relation in Eq. \ref{46} is indeed 
satisfied.  Therefore, the reaction $\gamma p\to K^0\Sigma^+$ could
be regarded as a mirror of the reaction $\gamma n\to K^+\Sigma^-$
in the isospin space.  The similar relations are also true for the reactions 
$\gamma p\to K^+\Lambda$ and $\gamma n\to K^0\Lambda$,
 $\gamma p\to K^+\Sigma^0$
and $\gamma n\to K^0\Sigma^+$,  $\gamma p \to \pi^+ n$ and 
$\gamma n\to \pi^- p$, and $\gamma p\to \pi^0 p$ and $\gamma n\to \pi^0 n$,
in which the relation in Eq. \ref{46} is satisfied\footnote{Except the states
$(70,^4N)$, of which the photon transition amplitudes vanish for the proton
 target}.   Thus,  the coefficient $g_R$ for the processes with the neutron
 target can be deduced from that of the proton target according to their 
isospin couplings,  and this result seems to be more general than the 
$SU(6)\otimes O(3)$ basis used here.  This also gives us an important 
constraint in predicting the reaction of neutron targets from the 
proton target results.  

If one intends to calculate the reaction beyond 2 GeV in the center of mass 
frame, the higher resonances with quantum number $N=3$ and $N=4$ must be 
included.  There are only a few resonances around 2 GeV that can be 
in principle classified as $N=3$ resonances, in particular the resonances 
$S_{31}(1900)$ and $D_{35}(1930)$.   However, we do not expect these 
resonances contribute significantly since they are lower partial wave 
resonances. Instead, we adopt an approach that treats the resonances 
for $N\ge 2$ as degenerate, the sum of the transition 
amplitudes from these resonances can be obtained through the approach in 
Ref. \cite{zpli93}. in Eqs. \ref{51} and \ref{52} becomes 
in the s-channel.  Generally, the resonances with
larger quantum number $N$ become important as the energy increases.
Note that the amplitude ${\cal O}_n^2$ generally differs from
 the amplitude ${\cal O}_n^3$ by a factor of 
$\left (-\frac 12\right )^n$,  this shows that 
the process that the incoming photon and outgoing meson are
 absorbed and emitted by the same quark becomes more and more dominant
as the energy increases.  Furthermore, 
the resonances with partial wave $L=N$ become dominant, of which the 
isospin is 1/2 for $\gamma N\to K \Lambda$ and $\gamma N\to \eta N$
 and 3/2 for $\gamma N\to K\Sigma$ and $\gamma N\to \pi N$.  Thus, 
we could use the mass and decay width of the high spin 
states in Eq. \ref{43}; the resonance $G_{17}(2190)$ for the $n=3$ 
states and the resonance $H_{19}(2220)$ for the $n=4$ states
in $\gamma N\to K \Lambda$ and $\gamma N\to \eta N$, and the 
resonance $G_{37}(2200)$ for the
$n=3$ states and the resonance $H_{3,11}(2420)$ for n=4 states in $\gamma 
p\to K\Sigma$.  Indeed, only the couplings for the high spin states
are strong enough to be seen experimentally, and this is consistent with the 
conclusions of the quark model.

\subsection*{4 Discussions}

Eq. \ref{44} establishes the connection between the transition 
amplitudes of the s-channel resonances and their underlying spin flavor
structure.  The relative strength and phase for each s-channel
resonance are determined by the $SU(6)\otimes O(3)$ symmetry 
so that no additional parameters are required.  Therefore, there are
some important features of the s-channel resonances in meson photoproductions
that can be discussed without numerical evaluation.  
Here we highlight some of them.

First, the S-wave resonances play very important roles 
in the threshold region, of which the transition amplitudes
are determined by $E^+_0$ transition.  This is particular true for the
kaon and $\eta$ production, in which masses 
of these S-wave resonances are sandwiched between their
 threshold energies.  Moreover, the effects of the S-wave resonances
are enhanced for the neutral meson production, since the seagull term
that dominates in this region does not contribute.
This has been widely recognized in the $\eta$ 
photoproduction\cite{zpli952,muko}, and their 
contributions to the threshold pion-photoproductions have been discussed 
recently\cite{zpli94}.  The same is true for the kaon-photoproductions 
as well.  Therefore, the kaon and $\eta$ productions in the threshold 
region provides very important probe to the structure of these s-wave
resonances.   In Ref. \cite{zpli961}, we showed that the kaon productions
experiments may provide us information on the existence of a quasi bound
$K\Lambda$ or $K\Sigma$ state, which has the same quantum number as the
the resonance $S_{11}$.  This will help us to understand the puzzle 
that the  decay into $\eta N$ is enhanced for the 
the resonance $S_{11}(1535)$ and suppressed for 
the resonance $S_{11}(1650)$.   

Second, assuming the $\eta^\prime$ quark coupling is either pseudo scalar
or pseudo vector,  one could extend this approach from $\eta$ 
to $\eta^\prime$ photoproduction.  An interesting prediction\cite{etap}
 from the quark model emerges for the $\eta'$ photoproduction; 
the threshold behavior of the $\eta'$ photoproduction is dominated by the 
off-shell contributions from the s-wave resonances in the second resonance 
region,  which can be tested in the future CEBAF experiments\cite{rechi}.
 This can be understood by the relative  strength of the CGLN 
amplitudes  between the s-wave resonances
in the second resonance region and the resonances around 2.0 GeV in 
the quark model.  There are two $S_{11}$ resonances with isospin 1/2
in the second resonance region.  The CGLN amplitudes for these two resonances
are proportional to that in Eq. \ref{512}, in which the leading term 
does not depend on the outgoing meson momentum ${\bf q}$.
On the other hand, the $S$ or $D$ wave resonances around 2 GeV 
belong to $n=3$ in the harmonic oscillator basis.  Acording to Eqs. \ref{51} 
and \ref{52}, the amplitudes for the $S$ and $D$ wave 
resonances with $n=3$ are at least proportional to ${\bf q}^2$ comparing
to the ${\bf q}$ dependence of the amplitude of the $S_{11}(1535)$
in Eq. \ref{512}, as the wavefunctions for the $S$ and $D$ wave resonances
with $n=3$ are orthogonal to that of $S_{11}$ resonances in the second
resonance region.  This leads to a smaller contributions from the
$S$ and $D$ wave resonances around 2 GeV to the threshold region of the
$\eta^\prime$ photoproductions.

Finally, the higher partial wave resonances become more important
as the energy increases.  Notice that
the CGLN amplitudes for the P-wave resonances with $N=2$, such as 
$P_{11}(1440)$ and $P_{11}(1710)$, are much smaller 
than those for the resonances $F_{15}(1680)$ and $F_{37}(1950)$.  
For the processes $\gamma N\to K\Sigma$, the 
contributions from the isospin 3/2 states, in particular those resonances 
in ${\bf 56}$ multiplet, are dominant.  Therefore, the processes $\gamma 
p\to K^+\Sigma^0$ and $\gamma p\to K^0\Sigma^+$  provide us a very 
important probe to the resonances with isospin 3/2, a particular example is 
the resonances $F_{37}(1950)$, $F_{35}(1905)$, $P_{33}(1920)$ and 
$P_{31}(1910)$.  It should be pointed out that the F-wave resonances
with isospin 3/2 were not included in most investigations.  It raises
the question whether these calculations are reliable beyond the 
threshold region.

\subsection*{5. Conclusion}

A comprehensive and unified appraoch to the pseudo-scalar meson 
photoproductions is presented in this paper. The quark model approach 
represents a significant advance in the theory of the meson 
photoproductions. It introduces the quark and gluon degrees of freedom
 explicitly,  which is an important step towards establishing the 
connection between  the QCD and the reaction mechanism.  It highlights 
the dynamic roles by the s-channel resonances,  in particular the roles 
of the $S_{11}$ resonances in the threshold region of the $K$, $\eta$ 
and $\eta^\prime$  photoproductions. 

Moreover, it should be pointed out that Eqs. \ref{231} and \ref{232} 
in the u-channel and Eqs. \ref{241} and \ref{242} in the s-channel 
are more general.
They correspond to the pseudo scalar meson photoproductions at the quark 
level, which are independent of the final states.   Thus, if we replace the
final nucleon and the $\Sigma$ states by $\Delta(1232)$ and $\Sigma^*$ 
states, the formulism presented here could be extended to the reactions
$\gamma N \to \pi \Delta(1232)$ and $\gamma N\to K^+\Sigma^*$, which
 have not been investigated in the literature.  

The initial results of our calculations in the $\eta$\cite{zpli952} 
and $K$\cite{zpli96} photoproductions has shown very good agreements 
between the theory and  experimental data  with far less parameters. 
The challenge for this approach would be to go one step further so that the 
quantitative descriptions of meson photoproductions, in particular 
the polarization observables that are sensitive to the detail structure
of the s-channel resonances, could be provided.  The numerical evaluations
of the $\pi$ and $K$ photoproductions in this approach are currently 
in pregress, which will be reported elsewhere.  
 
\subsection*{Acknowledgment}
The financial support from Peking University is gratefully acknowledged.

\subsection*{Appendix}

The matrix element for the nucleon pole term in the s-channel
is found to be
\begin{eqnarray}\label{70}
{\cal M}_N=\omega_m e^{-\frac {{\bf q}^2+{\bf k}^2}{6\alpha^2}}
\left ( \fs+\fn\right )\left (e_N
-\frac {{\bf k}^2}{4P_N\cdot k}\mu_N \right )   
 \vsig\cdot\vep \nonumber \\  +  ie^{-\frac {{\bf k}^2+{\bf q}^2}
{6\alpha^2}}\left [ \frac {\omega_m}2\left (\fs+\fn\right)+1\right ]\frac 
{\mu_N}{2P_N\cdot k} \vsig 
\cdot {\bf q} \vsig \cdot (\vep \times {\bf k}) \nonumber \\ +
  e^{-\frac {{\bf k}^2+{\bf q}^2}{6\alpha^2}} \left (
\fs+\fn \right ) \frac {e_N\omega_m}{4P_N\cdot k}\vsig \cdot {\bf k} 
\vep\cdot {\bf q} \nonumber \\ +
e^{-\frac {{\bf k}^2+{\bf q}^2}{6\alpha^2}} \left
[ \frac {\omega_m}2\left (\fs+\fn\right)+1\right ]\frac
{e_N}{2P_N\cdot k} \vsig\cdot {\bf q} \vep\cdot {\bf q}
\end{eqnarray}
where $P_N\cdot k=\omega_{\gamma}(E_N+\omega_{\gamma})$,
  $\mu_N$ is the magnetic moments of the nucleon, $e_N$ is the total
charge of the nucleon.

The matrix elements for 
the U-channel $\Lambda$ and $\Sigma$ exchange term in the kaon production
is 
\begin{eqnarray}\label{71}
{\cal M}_{\Lambda\Sigma}=-e^{-\frac {{\bf k}^2+{\bf q}^2}{6\alpha^2}}
\frac {M_f}{2M_N}\left (\frac {\mu_f}{P_f\cdot k}+\frac 
{g_{\Lambda\Sigma}\mu_{\Lambda\Sigma}}{P_S\cdot k\pm \delta m^2}\right ) 
\nonumber \\ 
\bigg \{ \frac {\omega_m{\bf k}^2}2 \left ( \fs+\fn\right ) \vsig \cdot 
\vep +\nonumber \\ i \left [ \frac {\omega_m}2 \left ( \fs+\fn\right )
+1\right ] \vsig \cdot 
(\vep\times {\bf k}) \vsig \cdot {\bf q}\bigg \} \nonumber \\ -
 e^{-\frac {{\bf k}^2+{\bf q}^2}{6\alpha^2}} \left (
\fs+\fn \right ) \frac {e_f\omega_m}{4P_f\cdot k}\vsig \cdot {\bf k} 
\vep\cdot {\bf q} \nonumber \\ -
e^{-\frac {{\bf k}^2+{\bf q}^2}{6\alpha^2}} \left
[ \frac {\omega_m}2\left (\fs+\fn\right)+1\right ]\frac
{e_f}{2P_f\cdot k} \vsig\cdot {\bf q} \vep\cdot {\bf q}
\end{eqnarray}
where 
\begin{equation}\label{72}
g_{\Lambda\Sigma}=\left \{ \begin{array}{r@{\quad}l} 
\frac {g_\Sigma}{g_\Lambda} & \mbox {for $\gamma N\to K\Lambda$} \\
\frac {g_\Lambda}{g_\Sigma} & \mbox {for $\gamma N\to K\Sigma^0$} \\
0 & \mbox{ other processes} \end{array} \right. 
\end{equation}
is the ratio between the coupling constants for $\Lambda$ and $\Sigma$ 
final states, $\mu_{\Lambda\Sigma}=1.61$ is the magnetic moments for the 
transition between the $\Lambda$ and $\Sigma^0$ states, and $P_f\cdot k= 
E_f\omega_{\gamma}+{\bf k}\cdot {\bf q}$, notice that the final baryon 
state has the total momentum $-{\bf q}$ in the center of mass system.

The u-channel nucleon exchange for the $\eta$ and $\pi$ productions is 
\begin{eqnarray}\label{73}
{\cal M}_{u}=-e^{\frac {{\bf k}^2+{\bf q}^2}{6\alpha^2}}
\frac {\mu_f}{2P_f\cdot k} 
\bigg \{ \frac {\omega_m{\bf k}^2}2 \left ( \fs+\fn\right ) \vsig \cdot 
\vep +\nonumber \\ i \left [ \frac {\omega_m}2 \left ( \fs+\fn\right )
+1\right ] \vsig \cdot 
(\vep\times {\bf k}) \vsig \cdot {\bf q}\bigg \} \nonumber \\ -
 e^{-\frac {{\bf k}^2+{\bf q}^2}{6\alpha^2}} \left (
\fs+\fn \right ) \frac {e_f\omega_m}{4P_f\cdot k}\vsig \cdot {\bf k} 
\vep\cdot {\bf q} \nonumber \\ -
e^{-\frac {{\bf k}^2+{\bf q}^2}{6\alpha^2}} \left
[ \frac {\omega_m}2\left (\fs+\fn\right)+1\right ]\frac
{e_f}{2P_f\cdot k} \vsig\cdot {\bf q} \vep\cdot {\bf q}
\end{eqnarray}

The matrix element for the t-channel is
\begin{equation}\label{74}
{\cal M}_t=e^{-\frac {({\bf k}-{\bf 
q})^2}{6\alpha^2}}\frac {e_m(M_f+M_N){\bf q}\cdot \vep}{q\cdot k}\left 
(\fs\vsig \cdot {\bf q}-\fn \vsig \cdot {\bf k}\right )
\end{equation}

\newpage

\begin{tabular}{lcccccccc}
\multicolumn{9}{l}
{ Table 1:  The $g$-factors in the u-channel 
amplitudes in Eqs. } \\
\multicolumn{9}{l}
{\ref{32} and \ref{33} 
for different production processes.}\\[1ex]
\hline\hline
Reaction & & $g^u_3$ &  $g^u_2$ &  $g_v$ & $g_v^\prime$& 
$g_a^\prime$ &  $g_A$ & $g_S$\\[1ex] \hline
$\gamma p\to K^+\Lambda$  & &  -$\frac 13$ &  $\frac 13$ & 1 & 1 & 
1& $\sqrt{\frac 32}$ & -$\frac {\mu_{\Lambda}}3$ \\[1ex]
$\gamma n\to K^0\Lambda$ & & -$\frac 13$ & $\frac 13$ & 1 & -1 &
-1&  $\sqrt{\frac 32}$ & $\frac {\mu_{\Lambda}} 3$\\[1ex]
$\gamma p \to K^+\Sigma^0$ & & -$\frac 13$ & $\frac 13$ & -3 &  -7 & 
 9 & -$\frac 1{3\sqrt{2}}$ & $\mu_{\Sigma^0}$\\[1ex]
$\gamma n\to K^0\Sigma^0$ & & -$\frac 13$ & $\frac 13$ & -3 & 11 &
 -9& $\frac 1{3\sqrt{2}}$ &  $\mu_{\Sigma^0}$\\[1ex]
$\gamma p\to K^0\Sigma^+$ & & -$\frac 13$ & $\frac 43$ & -3 & 2 & 
 0 & $\frac 13$ &  $\frac {2\mu_{\Sigma^+}}{3}$ \\[1ex]
$\gamma n\to K^+\Sigma^-$ & & -$\frac 13$& -$\frac 23$ & -3 & 2 &
 0&  -$\frac 13$ &  0\\[1ex]
$\gamma p\to \eta p$ & & 1 & 0 & 1 & 0 & 0&  1 & 0 \\[1ex]
$\gamma n\to \eta n$ & &  -$\frac 23$  & $\frac 23$ & 0 & -1
 &  0&  1 & 0 \\[1ex]
$\gamma p\to \pi^+ n$ & & -$\frac 13$  & $\frac 13$ & $\frac 35$ & 
$\frac 15$ & $\frac 95$ & $\frac 53$ & -$\frac{2\mu_n}5$ \\[1ex]
$\gamma n\to \pi^- p$ & & $\frac 23$ & $\frac 13$ & $\frac 35$ 
&$\frac 15$ &- $\frac 95$  &  - $\frac 53$ & -$\frac{4\mu_p}{15}$\\[1ex]
$\gamma p\to \pi^0 p$ & & $\frac 7{15}$ & $\frac 8{15}$ & 
$\frac {15}7$ &  2 & 0 &  $\frac 5{3\sqrt{2}}$ & $\frac{8\mu_p}{15}$\\[1ex]
$\gamma n\to \pi^0 n$ & & -$\frac 2{15}$ & $\frac 2{15}$ & 6 &  -7 
& 0 & -$\frac 5{3\sqrt{2}}$ & $\frac{4\mu_n}5$\\[1ex]
\hline
\end{tabular}

\vspace{1.5cm}

\begin{tabular}{lcl}
\multicolumn{3}{l}
{Tbale 2: Meson transition amplitudes $A$ in the simple}\\
\multicolumn{3}{l}{harmonic oscillator basis.}\\[1ex]
\hline\hline
$(N,L)$ & Partial Waves &  $A$ \\ \hline
$(0,0)$ &  P &  $-\left (\frac {\omega_m}{E_f+M_f}+1\right )$\\[1ex]
$(1,1)$ &  S &  $\frac {\omega_m}{\mu_q}-\left
(\frac {\omega_m}{E_f+M_f}+1\right 
)\frac {2{\bf q}^2}{3\alpha^2}$\\[1ex]
$(1,1)$ & D & $-\left (\frac {\omega_m}{E_f+M_f}+1\right )$\\[1ex]
$(2,0)$ & P & $\frac {\omega_m}{\mu_q}-\left 
(\frac {\omega_m}{E_f+M_f}+1\right 
)\frac {{\bf q}^2}{\alpha^2}$\\[1ex]
$(2,2)$ & P & $\frac {\omega_m}{\mu_q}-\left 
(\frac {\omega_m}{E_f+M_f}+1\right 
)\frac {2{\bf q}^2}{5\alpha^2}$\\[1ex]
$(2,2)$ & F & $-\left (\frac {\omega_m}{E_f+M_f}+1\right )\frac {{\bf 
q}^2}{\alpha^2}$\\[1ex] \hline
\end{tabular}

\vfill

\newpage

\begin{table}{ Table 3: The CGLN amplitudes for the S-channel baryons 
resonances for the proton target in the $SU(6)\otimes O(3)$ symmetry limit,
where $k=|{\bf k}|$, $q=|{\bf q}|$, 
and $x=\frac {{\bf k\cdot q}}{kq}$. The CGLN amplitudes 
for the $N(^4P_M)$, $N(^4S_M)$, and $N(^4D_M)$ states are zero due to the 
Moorhouse selection rule, see text.}\\[1ex]
\begin{tabular}{cllll}\hline\hline
States & $f_1$ & $f_2$  & $f_3$ & $f_4$ \\[1ex]\hline
$\Delta(^4S_s){\frac 32}^+$ & $3\frac {kqx}{2m_q}$ & $2\frac 1 {2m_q}$ & 
$-3\frac 1{2m_q}$ & 0 \\[1ex]
$N(^2P_M){\frac 12}^-$ & $ \frac {\omega_{\gamma}}{12}\left (1+\frac k
{2m_q}\right ) $ & 0 & 0 & 0\\[1ex]
$N(^2P_M){\frac 32}^-$ & $-\frac {\omega_{\gamma}}{18}\left (1+\frac k
{2m_q}\right )\frac {q^2}{\alpha^2}$ & $-\frac {kqx}
{12m_q\alpha^2}$  & 0 & $\frac {\omega_{\gamma}}{6\alpha^2}$ \\[1ex]
$\Delta(^2P_M){\frac 12}^-$ & $ \frac {\omega_{\gamma}}{6}\left (1-\frac 
 k{6m_q}\right ) $ & 0 & 0 & 0\\[1ex]
$\Delta(^2P_M){\frac 32}^-$ & $-\frac {\omega_{\gamma}}{9}\left 
(1-\frac {k}{6m_q}\right )\frac {q^2}{\alpha^2}$ 
& $\frac {kqx}
{18m_q\alpha^2}$  & 0 & $\frac {\omega_{\gamma}}{3\alpha^2}$ \\[1ex]
$N(^2S_s^\prime){\frac 12}^+$ & 0 & $-\frac {k^2}{216m_q\alpha^2}$ & 
0 & 0 \\[1ex]
$\Delta(^4S_s^\prime){\frac 32}^+$ & $\frac {k^3qx}{36m_q\alpha^2}$
& $\frac {k^2}{54m_q\alpha^2}$ & 
$-\frac {k^2}{36m_q\alpha^2}$\\[1ex]
$N(^2D_s){\frac 32}^+$ & $\frac {k^2qx}{36\alpha^2}
\left (1+\frac {k}{2m_q}\right )$ & $\frac {
k^2}{216m_q\alpha^2}$ & $\frac {\omega_{\gamma}}{36\alpha^2}$ & 0 \\[1ex]
$N(^2D_s){\frac 52}^+$ & $-\frac {k^2qx
}{180\alpha^2}\left (1+\frac {k}{2m_q}\right )$
& $-\frac {k^2}{144m_q\alpha^2}\left (x^2-\frac 15 
\right )$ & $-\frac {
k}{180\alpha^2}$ & $\frac {k^2x}{36q\alpha^2}$ \\[1ex]
$\Delta(^4D_s){\frac 12}^+$ & 0 & $-\frac { k^2}{108m_q\alpha^2}$ & 0 
& 0 \\[1ex]
$\Delta(^4D_s){\frac 32}^+$ & 0 &  $-\frac {k^2}{108m_q\alpha^2}$ & 
$\frac {k^2}{36m_q\alpha^2}$ & 0\\[1ex]
$\Delta(^4D_s){\frac 52}^+$ & $\frac {k^2}{126m_q\alpha^2}$ 
& $\frac {k^2}{126m_q\alpha^2}\left (x^2
-\frac 15\right )$ & 
$\frac {k^2}{210m_q\alpha^2}$ & 0\\[1ex]
$\Delta(^4D_s){\frac 72}^+$ & $\frac {k^2}
{12m_q\alpha^2}\left (x^2-\frac 37\right )$
 & $\frac {k^2 }{21m_q\alpha^2}\left (x^2
-\frac 15\right )$ & 
$\frac {k^2}{12m_q\alpha^2}\left (x^2
-\frac 17\right )$ & 0\\[1ex]
$N(^2S_M){\frac 12}^+$ & 0  &$-\frac {k^2}{432m_q\alpha^2}$ & 0 & 
0\\[1ex]
$\Delta(^2S_M){\frac 12}^+$ & 0 & $-\frac {k^2}{648m_q\alpha^2}$
& 0 & 0\\[1ex]
$N(^2D_M){\frac 32}^+$ & $\frac {k^2qx}{72\alpha^2}
\left (1+\frac {k}{2m_q}\right )$ & $\frac {
k^2}{432m_q\alpha^2}$ & $\frac {k}{72\alpha^2}$ & 0 \\[1ex]
$N(^2D_M){\frac 52}^+$ & $\frac {-k^2
qx}{360\alpha^2}\left (1+\frac {k}{2m_q}\right )
$ & $\frac {-k^2}{288m_q\alpha^2}\left (x^2-\frac 15 
\right )$ & $\frac {-
k}{360\alpha^2}$ & $\frac {k^2x}{72q\alpha^2}$ \\[1ex]
$\Delta(^2D_M){\frac 32}^+$ & $\frac {k^2qx}{36\alpha^2}
\left (1+\frac {k}{2m_q}\right )$ & $\frac {-
k^2}{648m_q\alpha^2}$ & $\frac {k}{36\alpha^2}$ & 0 \\[1ex]
$\Delta(^2D_M){\frac 52}^+$ & $\frac {-k^2
qx}{180\alpha^2}\left (1+\frac {k}{2m_q}\right )
$ & $\frac {k^2}{432m_q\alpha^2}\left (x^2-\frac 15 
\right )$ & $\frac {-
k}{180\alpha^2}$ & $\frac {k^2x}{36q\alpha^2}$ \\[1ex]
\hline
\end{tabular}
\end{table}

\begin{table}{ Table 4: The CGLN amplitudes for the S-channel baryons 
resonances for the neutron target in the $SU(6)\otimes O(3)$ symmetry limit,
where $k=|{\bf k}|$, $q=|{\bf q}|$, and $x=\frac {{\bf k\cdot q}}{kq}$.}\\[1ex]
\begin{tabular}{cllll}\hline\hline
States & $f_1$ & $f_2$  & $f_3$ & $f_4$ \\[1ex]\hline
$N(^2P_M){\frac 12}^-$ & $\frac {-\omega_{\gamma}}{12}\left (1+\frac k
{6m_q}\right ) $ & 0 & 0 & 0\\[1ex]
$N(^2P_M){\frac 32}^-$ & $\frac {\omega_{\gamma}}{18}\left (1+\frac k
{6m_q}\right )\frac {q^2}{\alpha^2}$ & $\frac {kqx}
{36m_q\alpha^2}$  & 0 & $\frac {-\omega_{\gamma}}{6\alpha^2}$ \\[1ex]
$N(^4P_M){\frac 12}^-$ & $\frac {-\omega_{\gamma}k}{36m_q}
$ & 0 & 0 & 0\\[1ex]
$N(^4P_M){\frac 32}^-$ & $\frac {-\omega_{\gamma}kq^2}{135m_q\alpha^2}
$ & $\frac {-kqx}{90m_q\alpha^2}$  & 0 & 0\\[1ex]
$N(^4P_M){\frac 52}^-$ & $\frac {-\omega_{\gamma}kq^2}{6m_q\alpha^2}\left 
(x^2-\frac 15\right )$ & $\frac {-kqx}{10m_q\alpha^2}$  & $\frac 
{kqx}{6m_q\alpha^2}$ & 0\\[1ex]
$N(^2S_s^\prime){\frac 12}^+$ & 0 & $\frac {k^2}{324m_q\alpha^2}$ & 
0 & 0 \\[1ex]
$N(^2D_s){\frac 32}^+$ & $\frac {-k^3qx}{108m_q\alpha^2}$ & $\frac {-
k^2}{324m_q\alpha^2}$ & 0 & 0 \\[1ex]
$N(^2D_s){\frac 52}^+$ & $\frac {k^3qx
}{540m_q\alpha^2}$ & $\frac {k^2}{216m_q\alpha^2}\left (x^2-\frac 15 
\right )$ & 0 & 0 \\[1ex]
$N(^2S_M){\frac 12}^+$ & 0  &$\frac {k^2}{1296m_q\alpha^2}$ & 0 & 
0\\[1ex]
$N(^2D_M){\frac 32}^+$ & $\frac {-k^2qx}{72\alpha^2}
\left (1+\frac {k}{6m_q}\right )$ & $\frac {-
k^2}{1296m_q\alpha^2}$ & $\frac {-k}{72\alpha^2}$ & 0 \\[1ex]
$N(^2D_M){\frac 52}^+$ & $\frac {k^2qx
}{360\alpha^2}\left (1+\frac {k}{6m_q}\right )$
& $\frac {k^2}{864m_q\alpha^2}\left (x^2-\frac 15 
\right )$ & $\frac {
k}{360\alpha^2}$ & $\frac {-k^2x}{72q\alpha^2}$ \\[1ex]
$N(^4S_M){\frac 32}^+$ & $\frac {-k^3qx}{216m_q\alpha^2}$ & 
$\frac {-k^2}{324m_q\alpha^2}$ & $\frac {k^2}{216m_q\alpha^2}$ & 0\\[1ex]
$N(^4D_M){\frac 12}^+$ & 0 & $\frac {-
k^2}{1944m_q\alpha^2}$ & 0 & 0 \\[1ex]
$N(^4D_M){\frac 32}^+$ & $\frac {k^3qx}{162m_q\alpha^2}$
& $\frac {7k^2}{1944m_q\alpha^2}$ & $\frac {-k^2}{216m_q\alpha^2}$ & 0 \\[1ex]
$N(^4D_M){\frac 52}^+$ & $\frac {-k^3qx}{756m_q\alpha^2}$
& $\frac {-k^2}{756m_q\alpha^2}\left (x^2-\frac 15 
\right )$ & $\frac {k^2}{1260m_q\alpha^2}$ & 0\\[1ex]
$N(^4D_M){\frac 72}^+$ & $\frac {-k^3q}{72m_q\alpha^2}\left (x^2-\frac 
37\right )$
& $\frac {-k^2}{126m_q\alpha^2}\left (x^2-\frac 15 
\right )$ & $\frac {k^2}{72m_q\alpha^2}\left (x^2-\frac 17\right )$ & 0\\[1ex]
\hline
\end{tabular}
\end{table}

\vspace{1.5cm}

\vfill
\newpage

\begin{tabular}{lccccc}
\multicolumn{6}{l}
{ Table 5:  The $g_R$-factors in the s-channel resonance
amplitudes for}\\
\multicolumn{6}{l} 
{different production processes.}\\
\hline\hline
Reaction & $(56,^2N)$ & $(56, ^4\Delta)$ & $(70, ^2N)$ & $(70, ^4N)$ &
$(70, ^2\Delta)$ \\ \hline
$\gamma p\to K^+\Lambda$  &  1 & 0 & 1 & 0 & 0\\[1ex]
$\gamma n\to K^0\Lambda$ & 1 & 0 & 1 & 0 & 0\\[1ex]
$\gamma p \to K^+\Sigma^0$ & 1 & $\frac 83$ & -1 & 0 & 1 \\[1ex]
$\gamma n\to K^0\Sigma^0$ & 1 & -$\frac 83$& -1  & -2 & -1\\[1ex]
$\gamma p\to K^0\Sigma^+$ & 1 & -$\frac 43$  & -1 & 0 & -$\frac 12$
\\[1ex]
$\gamma n\to K^+\Sigma^-$ & 1 & $\frac 43$ &-1 & -2 & $\frac 12$\\[1ex]
$\gamma p\to \eta p$ & 1 & 0 & 2 & 0 & 0\\[1ex]
$\gamma n\to \eta n$ & 1 & 0 & 2 & 1 & 0\\[1ex]
$\gamma p\to \pi^+ n$ & 1 &  $\frac 4{15}$  & $\frac 45$& 0 & $\frac 1{10}$
 \\[1ex]
$\gamma n\to \pi^- p$ & 1 &  -$\frac 4{15}$  &$\frac 45$ & -$\frac 15$ & 
-$\frac 1{10}$\\[1ex]
$\gamma p\to \pi^0 p$ & 1 &  -$\frac 8{15}$  & $\frac 45$& 0 & $\frac 1{10}$
 \\[1ex]
$\gamma n\to \pi^0 n$ & 1 &  $\frac 8{15}$  &$\frac 45$ &  -$\frac 15$ 
& -$\frac 1{10}$\\[1ex]
\hline
\end{tabular}
\vfill

\end{document}